\documentclass[amsmath,amssymb]{revtex4}
\usepackage{amsmath}
\usepackage{amssymb}
\usepackage{graphicx}
\usepackage{dcolumn}
\usepackage{bm}
\def\eqnn#1{Eq.~(\ref{eq:#1})}

\def\prl{{ \sl Phys. Rev. Lett. }}

\def\figno#1{Fig.~\ref{fig:#1}}

\def\vev#1{\langle#1\rangle}
\def\Im{{\rm \,Im}}
\def\ns{{\cal N}}
\def\dpsn{N_{\rm SN}}
\begin{document}
\title{
  Direct optical observations of surface thermal motions at
  sub-shot noise levels
}
\author{
  Takahisa Mitsui,\footnote{E--mail:~{\tt
        mitsui@hc.cc.keio.ac.jp}.}
  Kenichiro Aoki\footnote{E--mail:~{\tt ken@phys-h.keio.ac.jp}. %
    Supported in part by the Grant--in--Aid for Scientific
    Research from the Ministry of Education, Culture,  Sports, Science
    and Technology of Japan. 
  }%
}
\affiliation{Dept. of Physics, Hiyoshi, Keio University, 
  Yokohama 223--8521, Japan}
\begin{abstract}
    We measure spectral properties of surface thermal fluctuations of
    liquids, solids, complex fluids and biological matter using light
    scattering methods. The random thermal fluctuations are delineated
    from random noise at sub-shot noise levels. The principle behind
    this extraction, which is quite general and is not limited to
    surface measurements, is explained.  An optical lever is used to
    measure the spectrum of fluctuations in the inclinations of
    surfaces down to $\sim 10^{-17}\rm\,rad^2/Hz$ at $1\sim10\,\mu$W
    optical intensity, corresponding to $\sim 10^{-29}\,\rm m^2/\rm
    Hz$ in the vertical displacement, in the frequency range
    $1{\rm\,}\rm kHz\sim10\, MHz$.  The dynamical evolution of the
    surface properties is also investigated. The measurement requires
    only a short amount of time and is essentially passive, so that it
    can be applied to a wide variety of surfaces.
\end{abstract}

\vspace{3mm}

\maketitle
Thermal fluctuations are literally ubiquitous and exist for everything
we see. 
One manifestation is the classic
Brownian motion observed by Robert Brown more than a century ago.
Surfaces of all objects are fluctuating and it would be quite
interesting to see them directly.
However, the {\it atomic} fluctuations themselves are small and
have been directly seen only on few types of surfaces, such as
``ripplons'' on liquid surfaces\cite{ripplon,ripplonExp} and high
power interferometry measurements of surface fluctuations on
mirrors\cite{mirror}.
We have directly observed the power spectra (1\,kHz$\sim$10\,MHz) of
thermal fluctuations on various kinds of surfaces, presenting some
of their interesting cases in this work.  The measurements utilize a
novel method that determines the local inclinations of the surface
with an optical lever, down to orders of magnitude below shot noise
levels, separating the random signal from random noise. 
The optical intensity used in our experiments is relatively low,
$1\sim10\,\mu$W at the photodetectors, with $0.1\sim1$\,mW power light
applied to the samples. The latter is higher since the surface light
measurements capture reflected light and the reflectivity of the
surfaces we observed are not high.
The obtained power spectra are analyzed and are related to
the physical properties of materials. 
Thermal fluctuations are of interest not only for fundamental physics
but can also let us probe various properties of matter
non--invasively\cite{Denk}.  Furthermore, our method allows us to make
delicate low noise measurements of random motions when random noise is
the main limiting factor.

The main difficulty in directly measuring properties of thermal
fluctuations is that the fluctuations are small and random.  In any
measurement, the detectors also generate some random noise, from which
the signal needs to be extracted. Since the signals themselves are
random, simple averaging will not suffice to separate out the signal,
unless it is large.
A detector measurement $D_1=S+N_1$ consists of the desired signal $S$
and some other random equipment noise $N_1$, which is independent of
$S$.
To obtain interesting results from physics point of view, we compute
its power spectrum.  Denoting the Fourier transforms with tildes, if
we simply average over data, the power spectrum is obtained as,
$\vev{|\tilde D_1|^2} = \vev{|\tilde S|^2} + \vev{|\tilde N_1|^2} $.
There is no way to separate out the random noise, unless the signal is
relatively large, $\vev{|\tilde S|^2} \gg \vev{|\tilde N_1|^2}$ .
However, even a small random signal can be recovered by making a
statistically independent measurement of the same signal.  This can be
achieved by measuring the same signal at the same time, with another
physically independent detector.  Denoting this independent output as
$D_2=S+N_2$, 
\begin{equation}
    \label{eq:psd2}
    \vev{\overline{\tilde D_1}\tilde D_2}\rightarrow
    \vev{|\tilde S|^2}\qquad({\cal N}\rightarrow\infty)
    \qquad,
\end{equation}
averaging over the correlation of the two signals. Here $\ns$ is the
number of averagings.
The relative error in this method is statistical and its
size is $\sim1/\sqrt{\ns}$.  
Clearly, this theory of noise reduction is not limited to surface or
optical measurements.
The crucial requirement for this method
to work is that the random noise in the two measurements $D_1,D_2$ are
decorrelated. 

\begin{figure}[htbp]
    \centering
   \includegraphics[width=8.5cm,clip=true]{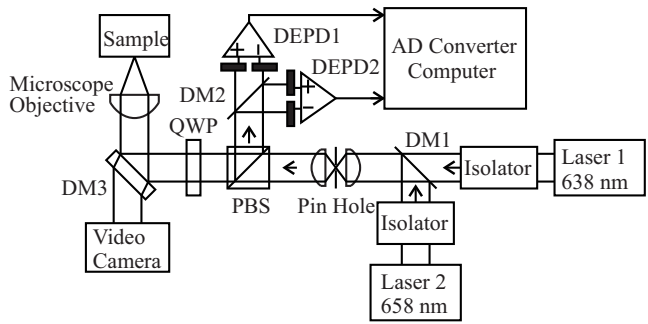}
   \caption{Experimental setup: Two independent laser beams are shone
     on the sample and their reflections are detected by dual-element
     photodiodes, DEPD1,2.  A DEPD measures the inclination of the
     surface through the difference in the light intensities.  A
     polarizing beam splitter (PBS) and a quarter wave plate (QWP) are
     included to extract the light reflected back from the sample
     efficiently.  Dichroic mirrors (DM1,2) are used to combine or
     separate the light with the two different wavelengths. The data
     is processed by a computer via an analog-to-digital (AD)
     converter. }
    \label{fig:setup}
\end{figure}
The experimental setup is shown schematically in \figno{setup}. In our
spectral measurements using surface light scattering, the average
inclination of the surface within the beam spot is measured by using
the difference between the reflection signals from two adjacent
photodiodes. Since we need two independent measurements to obtain the
signal as in \eqnn{psd2}, we use two sets of two photodiodes, DEPD1,2. 
Laser diodes with wavelengths 638\,nm and 658\,nm with a power 40\,mW
each were used as light sources. Due to the losses from stabilization,
the power of the beam applied to the sample is 2\,mW each, at most.
When the sample is an organic matter, such as rubber, thermal effects
from the beams are non--negligible so that a neutral density filter is
used to further reduce the power of the beams. The most significant
experimental limitations to the sensitivity in the measurement are
various sources of cross-talk. This makes the experimental realization
non-trivial, even though the principle explained above is simple.
Cross-talk can arise from such diverse sources as the the AD
converter, non-linear elements in the optical elements and coupling
through the electromagnetic fields. In our experiments, we measured
the size of the cross-talk prior to the measurements and corrected for
it.

The major causes of noise in the experiment are the shot noise and the
thermal noise in the detectors, often called Nyquist noise. These
types of noise exist even in ideal experimental situations.
Our method, as explained above, can suppress both types of noise and
measurements can be made at orders of magnitude below the noise level,
as explicitly demonstrated below.  The thermal noise level in the
detectors is mostly much less than the shot noise level and at most,
of the same order in our experiments.  Thermal noise can also be
reduced, at least in principle, by cooling the detector while such
methods are not applicable for the shot noise, which is essentially
quantum in nature. 
The main error in our work is the theoretical limitation due to the
number of averagings, $\cal N$.  In theory, as well as in practice,
this places a limitation that if a frequency resolution $\Delta f$ is
required, time $\ns/\Delta f$ is necessary to make the measurement.
In our experiments, the time used for the measurements varied from 30
seconds to 30 minutes, longer times being used for weaker signals.
One might point out that to overcome the error due to the shot noise,
one can consider raising the power of the signal by increasing the
light beam power. However, this is not always possible since raising
the power can affect the sample itself, making it impossible to obtain
meaningful results. Furthermore, large power is prohibitive in
situations where non--invasiveness is required, such as
medical applications. Given the same power, our method, when
applicable, can extract far weaker signals than those obtainable by
other methods.

We first observe and analyze thermal fluctuations on simple liquid
surfaces.  Thermal fluctuations of surfaces appear as capillary waves
called ``ripplons'' and light scattering from them have been studied
for some time\cite{ripplon,ripplonExp}. The dispersion relation of
ripplons for a simple liquid has been derived
theoretically, including its magnitude as\cite{Levich,Bouchiat} 
\begin{equation}
    \label{eq:ripplon}
    P_{\rm R}(q,\omega)={k_{\rm B}T\over \pi\omega}
    {q\tau_0^2\over\rho}\Im\left[(1+s)^2+y-\sqrt{1+2s}\right]^{-1}    
\end{equation}
Here, $s\equiv -i\omega\tau_0$, $\tau_0\equiv \rho/(2\eta q^2)$,
$y\equiv \sigma\rho/(4\eta^2 q)$. $\rho,\sigma,\eta$ are the density,
the surface tension and the viscosity of the liquid. $q,\omega$ are
the wave number and the frequency of the capillary waves. $k_{\rm B}$
is the Boltzmann constant and $T$ is the temperature. 
Our method differs from previous surface light scattering measurements
in that we directly measure the spectrum of surface inclination
fluctuations.  Our observations correspond to the spectrum
\begin{equation}
    \label{eq:ripplonW}
    S_{\rm R}(\omega)=\int\!\!\!{d^2q\over(2\pi)^2}\,
    q^2e^{-b^2q^2/8}P_{\rm R}(q,\omega)\qquad.
\end{equation}
Here, $b$ is the diameter of the gaussian beam and the $q^2$ factor
arises from observing inclinations. 
The vertical displacement spectrum is $b^2S_{\rm R}(\omega)$ so that
the measured spectral density in these fluctuations go down to
$\sim10^{-29}\rm\,m^2$/Hz.  When integrated over all frequencies, the
total vertical displacement is few angstroms.

\begin{figure}[htbp]
    \centering
   \includegraphics[width=8.5cm,clip=true]{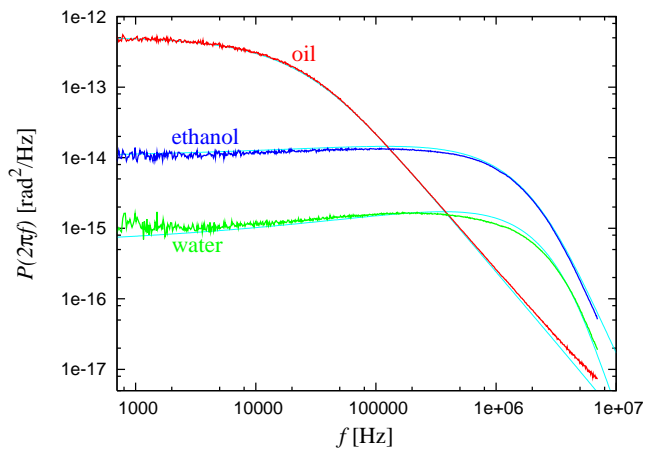}
   \caption{(Color online) Power spectrum of capillary waves on
     water(green), ethanol(blue) and oil surfaces(red). The
     theoretical predictions for them are also indicated (cyan) which
     agree quite well.}
    \label{fig:ripplon}
\end{figure}
The theoretical results \eqnn{ripplonW} agree quite well with the 
experimental observations, as shown in \figno{ripplon}.
The properties of the liquids are $(\rho\,{\rm [kg/m^3]},
\sigma\,{\rm [kg/s^{-2}]},\eta\,{\rm [kg\cdot
  m^{-1}s^{-1}]})=(1.0\times10^3,7.3\times10^{-2},1.0\times10^{-3})$,
$(0.79\times10^3,2.2\times10^{-2},1.1\times10^{-3})$,
$(0.92\times10^3,3.0\times10^{-2},0.124)$ for water, ethanol and
immersion oil. The beam diameters are $b=2.5\,\mu$m for oil, ethanol
and  $3.2\,\mu$m for water.
In \figno{ripplon}, the overall magnitude of the signal was adjusted
to fit the theoretical formula. This magnitude was independently
calibrated using a pizoelectrically driven mirror and was confirmed to
within a factor of two.  Due to the relatively high viscosity of oil,
there is a qualitative difference for the oil surface spectrum which
decays as $ f^{-2}$ for higher frequencies, when compared to those of
water and ethanol which decay as $ f^{-4}$. This is well reproduced
in the measurements.

The shot noise level
in our measurements is independent of $f$ and can be estimated as
$\dpsn(\omega)     = {N\!A^2e/(2I)}$, 
where $I$ is the photocurrent, $N\!A$ is the numerical aperture of the
objective lens and $e$ is the electron charge. We used a lens with
$N\!A=0.5$ throughout and the photocurrent in all our measurements is
$0.1$ to few\,$\mu$A.  The ambient temperature was 25$^\circ$\,C for
all our measurements.  
$\dpsn\simeq2\times10^{-15}\,\rm [rad^2/{\rm Hz}]$ in the
above ripplon measurements so that the measurements in \figno{ripplon} go
down to a couple of orders of magnitude below the shot noise level.

\begin{figure}[htbp]
    \centering
   \includegraphics[width=8.5cm,clip=true]{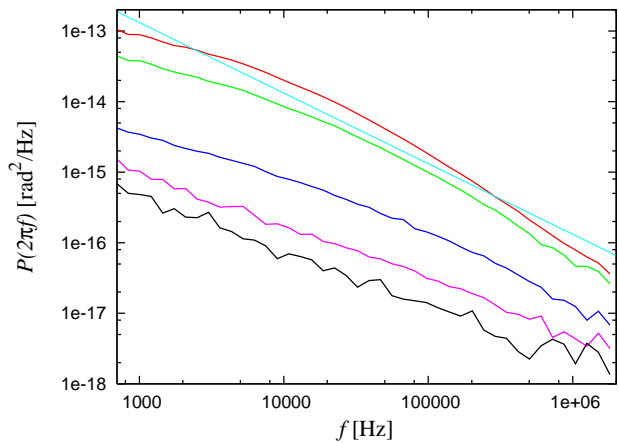}
   \caption{(Color online) Power spectrum of surface fluctuations of
     rubber in the 
     elongated direction, when the extension ratio is 1 (red),2
     (green), 3(blue), 4 
     (magenta), 6 (black).  The fluctuations decrease under
     stretching. For comparison,
     the theoretical formula corresponding to surface fluctuations of
     an elastic material 
     is also shown (cyan).
     $\dpsn\simeq1\times10^{-14}\rm\,[rad^2/ Hz]$ in this measurement.
   }
    \label{fig:rubber}
\end{figure}
Using the same method, the surface fluctuations of rubber with
varying strain are observed in \figno{rubber}.
An estimate for the fluctuation spectrum as an elastic
medium\cite{Saulson90,Levin,Braginsky99} is 
\begin{equation}
    \label{eq:elastic}
    S_{\rm S}(\omega)={16k_{\rm B}T\over \sqrt{\pi}\omega b^3 }
    {(1-\sigma_{\rm P}^2)\phi\over Y}\qquad.
\end{equation}
Here, $Y,\sigma_{\rm P},\phi$ are the Young's modulus, Poisson's ratio
and the loss angle of the material.  
In the
plot, we used natural rubber stretched to various lengths.  As a
guide, we indicated the spectrum in \eqnn{elastic} with 
frequency independent $Y=3.5\times10^{6}\,\rm [kg/(m\cdot s^2)],
\sigma_P=0.475, \phi=0.1$, which are typical values for rubber.  The
dependence on $f$ does not agree completely with this idealized
material, which is natural since $Y,\phi$ depend on $f$\cite{rubber}
and rubber is a complex material that changes its state as it is
elongated\cite{rubberStretch}.  Obviously, the dependence can be
perfectly reproduced if we assume a particular $f$ dependence for
$\phi/Y$.  The signal decreases as the Young's modulus increases when
the rubber is stretched\cite{rubberStretch}.  We have also measured
the fluctuations in the inclinations transverse to the direction of
extension and find them to be larger than those parallel to it, as
expected, indicating the existence of more flexibility in the
transverse direction.
The laser beam used had a diameter $1\,\mu$m with a power
150\,$\mu$W. Lowering this power did not change the spectrum.
On the other hand, if the power is raised substantially,
the beam affects the spectrum, presumably by
heating up and perhaps melting the material. Consequently, this result
is difficult to obtain without separating out the random noise using
the logic explained in \eqnn{psd2}.

\begin{figure}[htbp]
    \centering
   \includegraphics[width=8.5cm,clip=true]{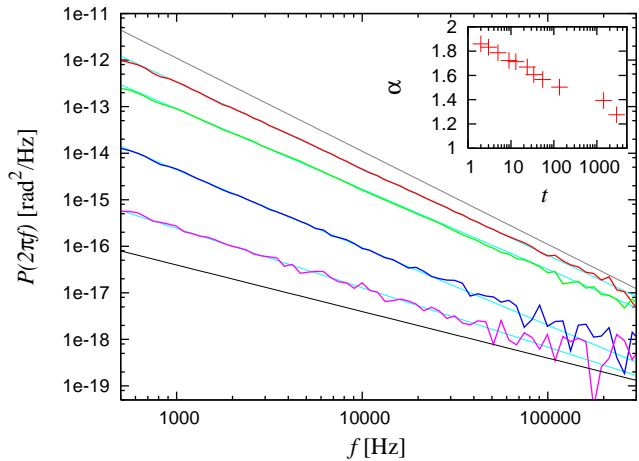}
   \caption{(Color online) Epoxy adhesive surface fluctuations 2
     (red), 9 (green), 24 
     (blue) minutes and 2 days (magenta) after application, together
     with fits to them to functions of the form const.$\times
     f^{-\alpha}$ (cyan), which work quite well. As a guide,
     theoretical results for a simple liquid (grey) and an elastic
     material (black) are shown also.  The fluctuations decrease with
     time. The change in $f$ dependence of the surface fluctuation
     power spectrum with respect to time $t$\,[min] shows a slow
     decrease between $\alpha=2$ and $\alpha=1$ (inset).
     $\dpsn\simeq4\times10^{-15}\rm\,[rad^2/Hz]$ in this measurement.
   }
    \label{fig:epoxy}
\end{figure}
We now consider a more complex material, an epoxy adhesive, whose
properties change over time, as the glue ``hardens''.  Our method
allows us to obtain spectral properties quickly without any contact
and we can see how the spectrum changes with time. The measurements
are shown in \figno{epoxy} where we used  a laser beam 
a diameter of $1\,\mu$m.
The thermal fluctuation spectrum at each instant can be well described
by a simple power dependence $P(f)\sim f^{-\alpha}$.  This power
$\alpha$ slowly decreases with time with values between two and one,
as in \figno{epoxy}~(inset). We recall that for highly viscous fluids,
$\alpha=2$ in this frequency range, as can be seen in \figno{ripplon},
and $\alpha=1$ for elastic materials, as in \eqnn{elastic}. The
results are quite consistent with an evolution of the epoxy adhesive
between these two states. For comparison, the ripplon spectrum in
\eqnn{ripplonW} with typical values for an epoxy adhesive,
$(\rho\,{\rm [kg/m^3]}, \sigma\,{\rm [kg/s^{-2}]},\eta\,{\rm [kg\cdot
  m^{-1}s^{-1}]})=(1.5\times10^3,4.0\times10^{-2},5000)$ and the
spectrum for an elastic material with $(1-\sigma_{\rm
  P})^2Y/\phi=1.5\times10^{11} \rm \, [kg/(m\cdot s^2)]$ are also
indicated.  Phenomenologically, the time dependence of $\alpha$ can be
well described by a logarithmic one in the relevant region (see
\figno{epoxy} inset).
The reason $\alpha$ decreases with
time in such  manner is worth further study.

\begin{figure}[htbp]
    \centering
   \includegraphics[width=8.5cm,clip=true]{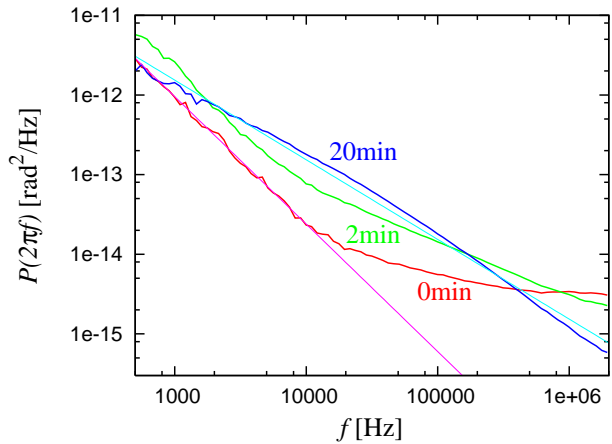}
    \caption{(Color online) Surface fluctuations of an ayu eye when it
      is wet (red) and 2 (green), 20 (blue) minutes
      thereafter. $6.0\times10^{-8}f^{-1.6}$ is shown (magenta). 
      Theoretical
      formula for an elastic material is also indicated (cyan). 
      $\dpsn\simeq2\times10^{-14}\rm\,[rad^2/Hz]$ in this measurement.
    }
    \label{fig:fisheye}
\end{figure}
Our method is well suited to biological materials since low power
laser beams can be used. \figno{fisheye} shows the measurements of the
surface fluctuation spectrum of an ayu (sweetfish) eye as it
dehydrates.  The laser beam in the measurement has a power of
200\,$\mu$W and a diameter $1\,\mu$m.  Measurements were performed
when the eye was wet and then after certain time had passed.
Not only quantitative, but clear qualitative differences in the
fluctuation spectrum can be seen with time.  
At the beginning, the eye surface is wet and at lower frequencies,
$f\lesssim40$\,kHz, the fluctuations decay as $\sim f^{-1.6}$ as can
be confirmed in \figno{fisheye}. This property is similar to that of a
highly viscous liquid like oil in \figno{ripplon} and viscous complex
fluid like epoxy in \figno{epoxy}. For higher frequencies,
$f\gtrsim40$\,kHz, the spectrum has almost no frequency dependence and
behaves similarly to the water spectrum in \figno{ripplon}, including its
magnitude. This suggests that the material contains water
substantially and is a gel like material.
With time, water evaporates and the thermal fluctuation spectrum
changes from that of a fluid to that of a solid.
We show the surface fluctuations of an elastic material in
\eqnn{elastic} with $(1-\sigma_{\rm P})^2Y/\phi=4.0\times10^6\,\rm
[kg/(m\cdot s^2)]$ in \figno{fisheye}, which describes the spectrum
after 20 minutes quite well. This spectrum is similar to the rubber
surface fluctuation spectrum in \figno{rubber}.

In this work, we used an optical lever\cite{opticalLever} to measure
power spectra of thermal fluctuations of a wide variety of surfaces,
from simple liquids to biological matter.
By analyzing the fluctuation spectra of various types of matter and
relating their spectra to their physical properties, the fluctuation
spectra of complex materials could be qualitatively explained  from the
understanding of the spectra of simpler matter.
The reason it is possible to make these sensitive measurements at our
low optical intensity is because the measurements were performed down to
orders of magnitude below the shot noise level.
There are situations, such as gravitational wave measurements, which
are usually believed to be shot noise limited\cite{GS95,gravWaveExp} and
our method can perhaps significantly improve the capabilities of those.
The measurements can be non--invasive and is applicable to all kinds
of surfaces, including biological matter and may be effective in
studying biological phenomena, such as the dry eye syndrome. 
Our measurement requires relatively a short time, allowing us to take
spectral snapshots of surfaces, observing the time evolution of
physical properties, as exemplified above.

\end{document}